\documentstyle[preprint,aps]{revtex}

\begin{document}
\draft
\preprint{Submitted to Physical Review Letters}
\title{Exact Results at the 2-D Percolation Point}
\author{P. Kleban$^1$ and R. M. Ziff,$^2$}
\address{$^1$
Laboratory for Surface Science and Technology
\& Department of Physics and Astronomy,
University of Maine,
Orono, ME 04469}
\address{$^2$ Department of Chemical Engineering,
University of Michigan, Ann Arbor, MI 48109-2136}

\date{\today}
\maketitle
\begin{abstract}

We derive exact expressions for the excess number of clusters $b$ and
the excess cumulants $b_n$ of a related quantity at the 2-D percolation point.
High-accuracy computer simulations are in accord with our predictions.  $b$ is a
finite-size correction to the Temperley-Lieb or Baxter-Temperley-Ashley
formula for the number of clusters per site $n_c$ in the infinite system
limit; the $b_n$ correct bulk cumulants. $b$ and $b_n$ are universal, and thus
depend only on the system's shape.  Higher-order corrections show no
apparent dependence on fractional powers of the system size.

\end{abstract}
\pacs{PACS numbers(s): 05.70.Jk, 11.25.Hf, 64.60.Ak}

\narrowtext
Percolation is perhaps the simplest non-trivial model in
statistical mechanics.  A broad array of techniques have been brought to
bear on it by researchers in a variety of disciplines, and it has continued
as an active research area for many years (for reviews, see \cite{Kest,SA}).  In
this work, we restrict ourselves to the case of two-dimensional systems at
the percolation point $p_c$.  Even this limited subject has been and continues
to be studied via renormalization group (\cite{AH} and references therein),
conformal field theory \cite{DF,Cardy,Watts}, Coulomb gas methods \cite{FSZ,Pinson},
computer simulation (\cite{Langlands,ZFA} and references therein), as an example of
supersymmetry \cite{Sal}, and using rigorous mathematical methods (see \cite{Kest} and
references below).

  	In this Letter we consider the average number of clusters
(connected graphs) $\langle N_C \rangle$ on a lattice of $N$ sites and in addition the
cumulants $C_n$ of a related quantity, $N_C + N_B/2$, where $N_B$ is the number of
bonds.  We derive new exact results for the excess or finite-size
correction term $b$ to the former and $b_n$ to the latter.  By use of conformal
field theory and Coulomb gas methods explicit expressions for both
quantities on a torus (including the cylinder limit), or a rectangle (with
conformally invariant boundary conditions) are determined.  These
expressions depend only on the aspect ratio of the geometry in question.
They are universal, and therefore apply to any particular realization, e.g.
lattice and percolation type.  For bond percolation on a square (triangular) lattice, the
formula for $n_c$, the number of clusters per site in the infinite system
limit, is known from the work of Temperley and Lieb \cite{TL}
(Baxter, Temperley and Ashley \cite{BTA}).  Thus $b$
represents the
 finite-size correction to
either of these results.  However, being universal, its validity is more
general and applies to any system at
criticality.  For the torus or cylinder, where $b$ is the leading finite-size
correction, our formulas agree extremely well with the results of computer
simulations.

Our work was motivated by the results of \cite{ZFA}, who found numerically
that $b$ for critical percolation in large systems appears to be a universal
quantity, dependent upon the aspect ratio $r$ but independent of the type of
lattice or percolation being considered.  Their values of $b(r)$ are plotted
in Fig.~1 below, while Table I presents new results for the infinite
aspect-ratio limit, where the torus becomes a cylinder and $b(r) \to \tilde b\, r$.
Here we find the identical value $\tilde b = 0.3608(1)$ for bond percolation on both
square and triangular lattices, which demonstrates the universality of this
quantity, and agrees with the conformal prediction to all figures.

Hu and Lin have studied the universality of a
related but different quantity, the number of percolating clusters, via
Monte Carlo simulation \cite{hu}.  Some recent theoretical results for
this quantity are given by Cardy \cite{cardy97}.

We also report (Table I) new results for a few excess cumulants $b_n$
of the quantity $N_C + N_B/2$, specifically $b_2$, which corrects the fluctuation,
$b_3$, which enters in the excess skewness and $b_4$, which is related to the
excess kurtosis.  For the square lattice, the bulk value of the fluctuation is
given by Temperley and Lieb \cite{TL}.  For percolation on a torus, we find
non-monotonic behavior for $b_2$ as a function of aspect ratio.  Good
agreement with computer results for $b_2$, $b_3$, and $b_4$ is obtained.

New computer simulations were carried out to investigate the
behavior of $\langle N_C \rangle$ and $C_n$, $n=2,3,4$,
in the cylindrical limit (high aspect ratio torus).
We used tori of dimensions $ L \times L'$ with $L'/L \ge 32$ and $L = 4, 5, 6, 7, 8, 
10, 12, 16,$ with additional runs at $L =32, 64, 128, 256$ where we measured $\langle N_C \rangle$
only.   About $10^{12}$ lattice points were generated for each lattice size,
requiring a total of several months of workstation computer time.  Such extensive work
was necessary to pinpoint the higher-order corrections to $b$ and the higher
cumulants.  With these aspect ratio $ \ge 32$, it was verified that no cluster
percolated the long way around the torus, so infinite cylinders were
effectively represented.  Most of our calculations were done for bond
percolation on a square lattice.  We also  measured $\langle N_C \rangle$  for bond
percolation on the triangular lattice, which can also be wrapped around to
make a simple cylinder, in order to check the universality hypothesis.  In
addition, new
simulations for $ L\times L$ systems were carried out.

	 Our results solidify the identification of critical percolation as
a realization of conformal field theory.  This connection was first noticed
some years ago \cite{DF} and made particularly explicit by Cardy \cite{Cardy}, who studied
the crossing probabilities via correlation functions of boundary operators.
Here we operate at a simpler level, employing the partition function.  Thus
our results may suggest a way to a deeper understanding of this
identification --- an independent derivation of $b$, for instance, would be very
interesting.  There is in fact  recent progress of this type for the crossing
probability in Voronoi percolation \cite{BS}.
	
At the critical point, the partition function of the $Q$-state Potts
model, for $Q$ real, may be expressed in random cluster form (see \cite{B})
\begin{equation}
Z = \sum_{\rm graphs} Q^{N_C + N_B/2}
\label{eq1}
\end{equation}
where the sum extends over all graphs (possible placements of $N_B$ bonds on
the edges of a lattice), $N_C$ is the number of clusters (connected
vertices, including isolated points), and the coupling has been set to
unity.  The transition is second-order for $0 \le Q \le 4$, with critical
temperature given by $e^{\beta_c} - 1 = Q^{1/2}$.
Here we make use of the $Q = 1$ case, where the
graphs are equally weighted, so that $Z$ describes percolation \cite{KF}.
Similar results may be obtained for the Ising ($Q = 2$) and other
universality classes; these will be reported elsewhere.

	The cumulants $C_n$ of $N_C + N_B/2$ follow immediately,
\begin{equation}
C_n = \left( Q {d \over dQ} \right)^n \ln Z
\label{eq2}
\end{equation}
Note that $C_1 = \langle N_C + N_B/2 \rangle $, $C_2$ gives the corresponding fluctuation, $C_3$
enters in the skewness, etc.

At the critical point, $Z$ is supposed to factorize
\begin{equation}
Z = \hat Z \cdot Z_u
\label{eq3}
\end{equation}
where $\hat Z$ is non-universal, depending on the lattice type, boundary
conditions, etc., while $Z_u$  encodes the universal information.  The
corresponding universal term $F_u = - \ln Z_u$  is the finite-size correction to the free
energy; it also generates the excess quantities studied here.  For a given
geometry, it depends only on the central charge
\begin{equation}
c = 1 - { 6 \left[ \cos^{-1} (\sqrt{Q}/2) \right]^2   \over  \pi  \left[\pi -  \cos^{-1} (\sqrt{Q}/2)
\right] }
\label{eq4}
\end{equation}
$\cite{FSZ}$ and the shape.

The simplest case is an infinite cylinder of width (circumference)
$L$.  Here, if we consider a length $L'$ of the cylinder,  $F_u$ is given by \cite{BCN,A}
\begin{equation}
F_u = - {\pi c \over 6} {L' \over L}
\label{eq5}
\end{equation}
For an infinite strip with edges (and conformally invariant boundary
conditions), the formula is the same, except that the 6 is replaced by 24
\cite{BCN}.

	Now for percolation ($Q = 1$, $c = 0$) on a cylinder (or torus) the
graphs are equally weighted so the number of bonds is exactly proportional
to the number of lattice sites.  Thus there is no finite-size correction to
$\langle N_B \rangle $.  The above then leads to
$C_1 - \langle N_B \rangle /2 = \langle N_C \rangle = n_c LL' +
\tilde b L'/L$ ($L'>>L$), with
\begin{equation}
\tilde b = {5 \sqrt 3 \over 24} = 0.360\,844\dots
\label{eq6}
\end{equation}
The agreement with computer simulation results is excellent, as shown in
Table I. \ For the strip (with edges), the r.h.s. of (\ref{eq6}) is divided by four.

For a torus, the universal factor $Z_u$   (for arbitrary $Q$) may be
evaluated via the work of di Francesco, Saleur and Zuber \cite{FSZ}, who employ a
Coulomb gas formulation.  This results in an expansion in powers of the
parameter $q = e^{-2 \pi r}$  , where $r = L'/L$ (length/width) of the torus.  For percolation,
the above procedure then leads to
$\langle N_C \rangle = n_c LL' + b$, with
\begin{eqnarray}
b = &&{5 \sqrt 3 \over 24} r + q^{5/4}\left( -{1\over2} + 2\sqrt{3} r \right)
+ q^{2}\left( -1 + \sqrt{3} r \right) \nonumber \\
&&+ \ldots + q^{5/48} + 2 q^{53/48} - q^{23/16} + q^{77/48} + \ldots, 
\label{eq7}
\end{eqnarray}
keeping all terms through $O(q^2)$.  The terms proportional to $r$ appear
because the powers of $q$ in $Z_u$  depend on $Q$, and thus contribute to the
derivative.  They include the leading result  (\ref{eq6})  for $r \to \infty$, which arises
from the derivative of the partition function of the unit operator.  Note
that the term correcting $Z_u$  for ``cross-topology" clusters is proportional to
$Q-1$ (see \cite{FSZ} (4.8)), and thus does not contribute this way, since $Q = 1$
here.  Fig.~1 compares  (\ref{eq7})  with computer simulations.  Excellent
agreement is found again.

By symmetry, $b$ is the same for $r$ and $1/r$, although this is not
evident from the few terms exhibited in (\ref{eq7}).  A graph of $b$ vs.~$r$ shows that
the leading behavior in the cylinder limit (the first term on the r.h.s.)
is corrected by a positive term that goes smoothly to zero as $r$ increases
from 1.  This term is such that $r = 1$ is the only minimum of $b$.

For a rectangle, with edges, of length $L$ and width $L'$, $Z_u$ is known (for arbitrary
$c$) via the results of Kleban and Vassileva \cite{KV}.  One can therefore
evaluate  $F_u$  for any $Q$ by the same procedure.  For percolation one has (here,
the finite-size correction to $\langle N_B \rangle$ is independent of $L'/L$, and is therefore
not included)
\begin{equation}
b = {5 \sqrt{3} \over 32 \pi} \ln L L' -  {5 \sqrt{3} \over 16 \pi} \ln [\eta(q)\eta(\tilde q)] ,
\label{eq8}
\end{equation}
where $\eta$ is the Dedekind $\eta$-function,
$q = e^{-2 \pi r}$   with $r = L'/L$, and $\tilde q = e^{-2 \pi /r}$.  However,
we have not attempted computer simulations in this case (or for the strip
with edges), since the leading finite-size correction will be due to a
non-universal edge term, making the precise determination of $b$ or $b_n$
difficult.  Note that the $Q$-dependence of $b_n$, for any $n$, involves the same
function of $q$ shown in (\ref{eq8}), by contrast to the torus.  This is related to
the fact that $Z$ on a rectangle generally couples only to the unit operator
\cite{KV}. Eq.~(\ref{eq8}) also assumes there is no $Q$-dependence of the matrix element or
constant term in $F_u$ (see \cite{KV} for details).

The fluctuation in $N_C + N_B/2$ is given by $C_2$.  For percolation on a
cylinder, its finite-size correction $\tilde b_2 L'/L$   follows from (\ref{eq5}).  We obtain
\begin{equation}
\tilde b_2 = {5 \sqrt{3} \over 36 } - {9 \over 16 \pi } = 0.061\,513\ldots
\label{eq9}
\end{equation}
Computer simulations results for this quantity are illustrated along with
(\ref{eq5}) in Table I.

The excess fluctuation of the number of clusters $N_C$
would also be of interest.  However, to calculate this requires the excess
for $\langle N_C N_B \rangle$, which is apparently not accessible.
 A similar situation occurs
for the leading fluctuations in the Temperley-Lieb theory \cite{ZFA,TL}.

	For $\tilde b_3$  and  $\tilde b_4$ we find, similarly from (\ref{eq2})--(\ref{eq5}), 
$\tilde b_3 = 5 \sqrt 3 / 36 - 9/8 \pi + 27 \sqrt 3 / 64 \pi^2 = -0.043\,500\ldots$   and  
$\tilde b_4 = 25 \sqrt 3 / 108 - 9/4 \pi + 27 \sqrt 3 / 16 \pi^2- 81 / 64 \pi^3 =  -.059\,933\ldots$.
Good agreement
with simulations is obtained here as well; see Table I. \ These values are
obtained by linear fits to the data vs.~$L^{-2}$, which also yields the bulk
values of $C_2$, $C_3$ and $C_4$.  Our results for $C_2$ and $C_3$ are in agreement with
the predictions $0.039\,445\ldots$ and $0.012\,913\ldots$ that follow from
\cite{TL}; we have not been able to obtain
$C_4$ either  analytically or numerically from that theory, but find from our simulations
$C_4 \approx 0.0028$.
Details will be presented elsewhere.  The $\tilde b_n$ results are less accurate than
those for $\tilde b$ because of the problem of extrapolating to infinity, and
because the error grows rapidly as $n$ increases (since higher moments are
involved).

The above also gives the excess fluctuations $b_2$ for percolation on a torus.  One finds
\begin{eqnarray}
b_2 &&= \tilde b_2 r + q^{5/48} \left( 1 - {11 \sqrt 3\over 48}r \right) - q^{5/24} \nonumber \\
&& + q^{53/48} \left( 2 - {11 \sqrt 3\over 24}r \right) - 4 q^{29/24} \nonumber \\
&& + q^{5/4} \left( {1 \over 2} + \left[ {53 \sqrt 3\over 24}
- {9 \over 4 \pi} \right] r + {3\over2}r^2\right) + \ldots
\label{eq10}
\end{eqnarray}
keeping terms through $O(q^{5/4})$ to show the appearance of $r^2$ in the higher
coefficients.  At $r = 1$, $b_2 = 0.105\,436\,634\ldots$ is at a local maximum
(note $b_2(r) = b_2(1/r)$). \ 
As the aspect ratio $r$
increases, $b_2$ decreases to a minimum of $0.103\,341\,600\ldots$ at
$r \approx 1.6177.$ \  As
$r$ increases further, $b_2$ increases, with limiting (large $r$) behavior
agreeing with (\ref{eq9}).  Such non-monotonic behavior with aspect ratio is very
reminiscent of the finite-size correction to the specific heat (which is
itself proportional to the excess of the fluctuation of $N_B$) in the Ising
model critical region.  This also exhibits non-monotonic behavior as a
function of $r$ \cite{FF} (see their Fig.~6).  This curious feature has been
explained in terms of a one-dimensional array of domain walls \cite{PK-GA}; a
connection to whatever mechanism is operating at the percolation point
would be very interesting.

We have also considered the convergence of the excess quantities.
This is equivalent to studying the higher order finite-size corrections
to the density or cumulants.
In all cases examined, the leading such term is proportional to $L^{-2}$;
however its coefficient is definitely not universal.  Fig.~2 illustrates
this for $b(r=1)$.  For the square lattice on
the cylinder, assuming the exact $n_c$ and $\tilde b$, an
analysis of the higher-order corrections shows no sign of a fractional
power of $L$, signifying an irrelevant singularity in the renormalization group,
as are seen in some other problems in percolation \cite{AH},
and suggests a simple analytic series
\begin{equation}
{\langle N_C \rangle \over L^2}  = n_c + {\tilde b \over L^2} + {0.180 \over L^4}
+ {0.69 \over L^6} + \ldots
\label{eq11}
\end{equation}
Eq.~(\ref{eq11}) fits all our measurements of $\langle Nc \rangle$ to nearly their full accuracy,
about $\pm 5\cdot10^{-7}$.  This form is consistent with corrections to
scaling due to the breakdown of rotational invariance on a square lattice
\cite{JLC-lH}.

	In summary, we have shown that at the percolation point in two
dimensions, the universality of the excess cluster number $b$, observed to
hold numerically in \cite{ZFA}, follows from conformal field theory.  We have
derived explicit expressions for $b$ and also for the excess cumulants of a
related quantity that are in complete agreement with simulation results.

We acknowledge useful conversations with J. L. Cardy, F. Y. Wu,
and thank C. Lorenz for obtaining numerical results for the triangular lattice.
RZ was supported  by the US 
National Science Foundation under Grant No.\thinspace DMR-9520700.

\begin{figure}
\caption{Excess number of clusters $b$ vs.\,aspect ratio $r$ at the
percolation point.  Solid line: exact results; points: simulations [10].
The simulations combine several types of percolation and lattice types.
The error is less than the size of the points. }
\end{figure}

\begin{figure}
\caption{$b_{\rm eff}(r=1)$ vs.\ $L^{-2}$. Points: simulation results for an $L \times L$ square
lattice, Solid line: fitting curve $0.88358 + 0.18 L^{-2}$ consistent with (11). $b_{\rm eff}$ is defined
as $\langle N_C \rangle - n_c L^2$, where $n_c = (3 \cdot 3^{1/2}  - 5)/2$ [10,12].
The conformal prediction for $b(r=1)$ from (7) is $0.883\,576\ldots$.  }
\end{figure}

\vfill\eject

\begin{table}
\caption{ Exact and simulated results for excess cumulants at the
percolation point.  See text for analytic expressions.
\label{table1}}
\begin{tabular}{lll}
Quantity & Exact&Simulation\\
\tableline
\tableline
$\tilde b$ &	0.360\,844...&	0.3608(1)\\
$ b_2(r=1)$ &$0.105\,437\ldots$&	$0.106 $\\
$\tilde b_2$ &$0.061\,513\ldots$&$	0.068$\\
$\tilde b_3$ &$-0.043\,500\ldots$&$	-0.041$\\
$\tilde b_4$ &$-0.059\,933\ldots$&$	-0.059 $\\
\end{tabular}
\end{table}

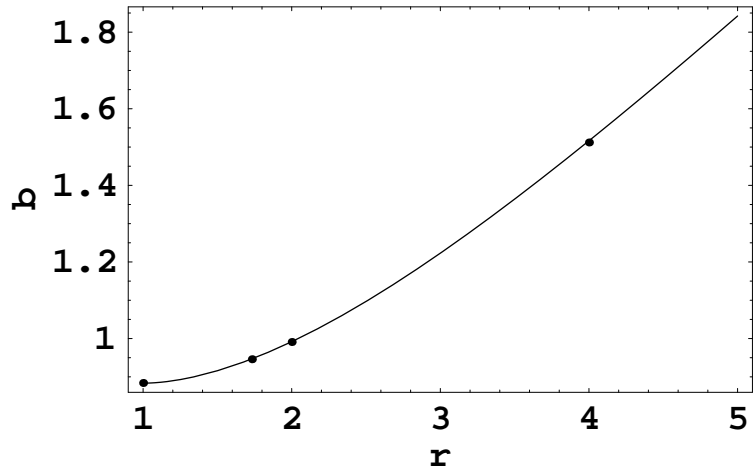
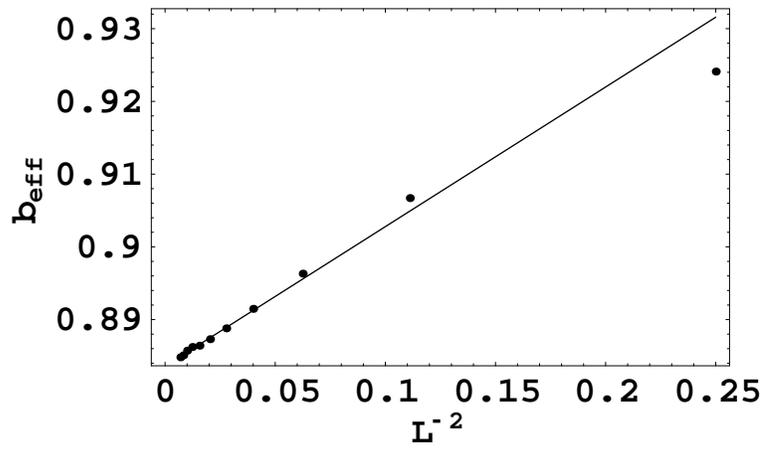
\end{document}